\documentclass[conference]{IEEEtran}
\IEEEoverridecommandlockouts
\usepackage{cite}
\usepackage{subcaption}
\ifCLASSINFOpdf
   \usepackage[pdftex]{graphicx}
   \DeclareGraphicsExtensions{.pdf,.jpeg,.png}
\else
  \usepackage[dvips]{graphicx}
  \DeclareGraphicsExtensions{.jpg}
\fi

\graphicspath{{./images/}} 

\usepackage{textcomp}
\usepackage{xcolor}

\usepackage{epstopdf}
\usepackage{amsthm}
\usepackage{color}
\usepackage{algorithm}
\usepackage{algorithmic}
\usepackage{enumerate}
\usepackage{enumitem}

\usepackage{etoolbox}  
\makeatletter
\patchcmd{\algorithmic}{\addtolength{\ALC@tlm}{\leftmargin} }{\addtolength{\ALC@tlm}{\leftmargin}}{}{}
\usepackage{amssymb}

\ifCLASSINFOpdf
   \usepackage[pdftex]{graphicx}
\else
\fi
\usepackage{cite}
\usepackage[cmex10]{amsmath}
\usepackage{array}
\usepackage{url}
\usepackage{makeidx}
\usepackage{verbatim}
\usepackage{multirow}
\usepackage{cases}

\newcommand{\beqn}{\begin{eqnarray}}
\newcommand{\eeqn}{\end{eqnarray}}
\newcommand{\bse}{\begin{subequations}}
\newcommand{\ese}{\end{subequations}}

\renewcommand\IEEEkeywordsname{Keywords} 

\usepackage{algorithm}
\usepackage{algorithmic}

\usepackage{etoolbox}  
\makeatletter
\patchcmd{\algorithmic}{\addtolength{\ALC@tlm}{\leftmargin} }{\addtolength{\ALC@tlm}{\leftmargin}}{}{}
\makeatother

\makeatletter 
\newcommand{\linebreakand}{%
  \end{@IEEEauthorhalign}
  \hfill\mbox{}\par
  \mbox{}\hfill\begin{@IEEEauthorhalign}
}
\makeatother 

\begin{document}

\title{Modeling Power Systems Dynamics with Symbolic Physics-Informed Neural Networks 
}
\author{Huynh T. T. Tran and Hieu T. Nguyen \\
{\it Department of Electrical \& Computer Engineering, North Carolina A\&T State University} \\

htran@aggies.ncat.edu, htnguyen1@ncat.edu}

\maketitle

\begin{abstract}

In recent years, scientific machine learning, particularly physic-informed neural networks (PINNs), has introduced new innovative methods to understanding the differential equations that describe power system dynamics, providing a more efficient alternative to traditional methods.
However, using a single neural network to capture patterns of all variables requires a large enough size of networks, leading to a long time of training and still high computational costs.
In this paper, we utilize the interfacing of PINNs with symbolic techniques to construct multiple single-output neural networks by taking the loss function apart and integrating it over the relevant domain.
Also, we reweigh the factors of the components in the loss function to improve the performance of the network for instability systems.
Our results show that the symbolic PINNs provide higher accuracy with significantly fewer parameters and faster training time.
By using the adaptive weight method, the symbolic PINNs can avoid the vanishing gradient problem and numerical instability.

\end{abstract}


\section{Introduction}
\label{intro}

Modeling the power system dynamics requires solving a set of complex nonlinear differential equations \cite{mohammadian2023gradient}.
This is a non-trivial task due to the large of components of the networks such as generators, loads, and transmission lines.
Traditional computers can use numerical discretization methods like Euler, the Runge-Kutta, and the backward differentiation formula to solve these differential equations.
However, the computational costs of these methods increase exponentially with system size \cite{kyriienko2021solving}.
In addition, with the growing integration of renewable energy sources, power systems have become more complex, creating new technical challenges \cite{xiao2022feasibility}.
Thus, the demand for modeling methods that enable high-accuracy modeling and lower power system costs has significantly increased \cite{mohammadian2023gradient}.

Scientific machine learning techniques, particularly combining machine learning with traditional scientific computing and mechanistic modeling \cite{zubov2021neuralpde}, can be considered as one of the alternative solutions for modeling the dynamics of complex systems such as the power grid. 
One of the key techniques in scientific machine learning is  Physics Informed Neural Networks (PINNs), i.e., combining deep neural networks (NN) with physics equations. 
Specifically, NN as a data-driven method can learn general solutions from a dataset and provides rapid approximation whereas the embedded physics equations 
help encourage consistency with the known physics of the system \cite{raissi2019physics}.
Thanks to the growth of available data and computer resources, PINNs have been applied to tackle several physics and engineering problems, such as modeling power system dynamics \cite{misyris2020physics}, heat transfer  \cite{cai2021physics}, fluid mechanics \cite{cai2021physics_fluid}, and high-speed aerodynamic flows \cite{mao2020physics}.


There have been several structures of PINNs proposed in the literature to model power system dynamics.
Ref. \cite{xiao2022feasibility} used the neural ordinary differential equations (ODEs) where the input variables are mapped to the derivative of the hidden variables and must first be integrated before their use \cite{chen2018neural}. 
Although neural ODEs can represent continuously defined time series dynamics with high accuracy, the hidden variables need to be integrated, which means they must be sent through an ODE solver.
Thus, the lack of knowledge of hidden variables causes difficulties in modeling the system \cite{kong2022dynamic}.
Another structure of PINNs had been used in \cite{mohammadian2023gradient, misyris2020physics, kong2022dynamic}, which are originally designed to solve partial differential equations (PDEs), a generalization of ODEs. 
By using an unsupervised strategy, they do not require labeled data derived from prior simulations or experiments whereas differential equations' solutions can be found by minimizing loss function optimization problems instead of directly solving governing equations \cite{cuomo2022scientific}.
However, this method requires all dependent variables to be defined on the full domain, which causes high computational cost \cite{zubov2021neuralpde}.
Also, its performance can suffer from kernel saturation issues if the NN is split into multiple outputs,

To handle these obstacles, we leverage the recent advance of science machine 
learning techniques provided in the ModelingToolkit (MTK) package \cite{ma2021modelingtoolkit} to construct a new structure of PINNs. Instead of a single large network, we split the network into multiple single-output NNs, in which each variable is represented by a network with a smaller size.
Additionally, we can integrate the loss function over the relevant domain for each portion of the loss function by using a symbolic interface,
therefore reducing the number of trainable parameters and the training time.
In certain scenarios, the varying scale of the loss function can lead to optimization challenges. 
To address this, we used the adaptive weight method, as outlined in \cite{wang2020understanding}. 
This approach involves re-evaluating the factors that contribute to the loss function and adjusting their weights to enhance the performance of the network.

The rest of the paper is organized as follows. 
Section \ref{PINNs} reviews the basic concept of physics-informed neural networks, introduces a new structure for NNs, and a method to reweigh the loss function components.
Section \ref{casestudy} represents the case studies and implementation.
Our numerical results are shown in Section \ref{Numerical} and Section \ref{conclu} concludes the paper.

\section{Physics-informed Neural Networks (PINNs)}
\label{PINNs}
In this section, we first review the basic concept of conventional PINNs and then provide the structure of symbolic PINNs and the adaptive weight method.

\subsection{Overview of conventional PINNs}
\label{overview}

\begin{figure}[t!]
    \centering
    \includegraphics[width = 0.42\textwidth]{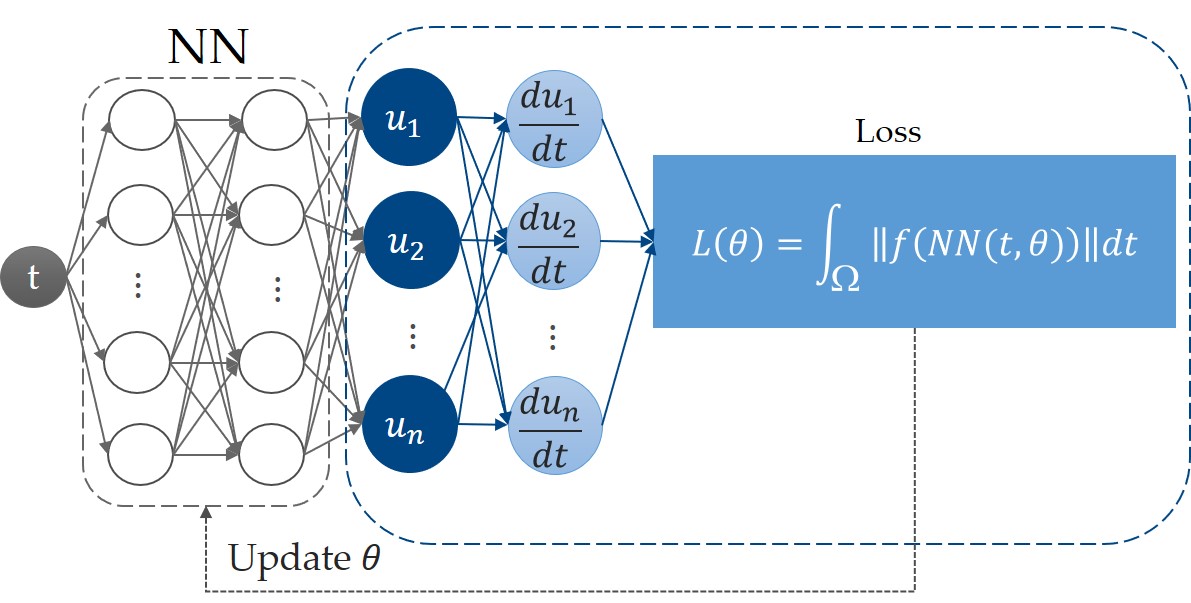}
    \caption{General structure of conventional PINNs \cite{cai2021physics}.}
    \label{fig: conventional_PINN}
    \vspace{-5pt}
\end{figure}
The structure of conventional PINNs is shown in Figure \ref{fig: conventional_PINN}.
In general, physical models for a system of partial differential equations (PDEs) can be defined as \cite{zubov2021neuralpde}:

\begin{equation}
    f(u,t) = 0, \quad \forall t \in \Omega,
    \label{general_DE}
\end{equation}
where $f$ is the residual of differential equations containing the nonlinear differential operators acting on $u(t)$, $u(t)$ are state variables, $t$ is the time (independent variable), and $\Omega$ is the computational domain. 
PINNs solve \eqref{general_DE} by using the Universal Approximation Theorem, in which if exist any sufficiently regular function $u(t)$, there is a large enough neural network $NN$ with the parameters $\theta$ (including weights and biases) such that $\| NN(t,\theta) - u(t) \| < \epsilon$ for all $t \in \Omega$. 
We can replace unknown $u(t)$ by a NN $NN(t,\theta)$ and find the parameters such that $f(NN(t,\theta)) \approx 0$ for all $t \in \Omega$.
Therefore, in PINNs, solving a system of differential equations is converted into an optimization problem with a loss function that is the sum of differences at every point within the domain and can be written as follows:
\begin{equation}
    \mathcal{L}(\theta) = \int_\Omega \| f(NN(t,\theta)) \|dt,
    \label{eq: 2}
\end{equation}
where $\| \cdot \|$ is the norm operator, and $\mathcal{L}(\theta)$ is the difference from the exact solution, and the goal is minimizing $\mathcal{L}(\theta)$.
If $\mathcal{L}(\theta) = 0$, then by definition, the outputs from the NN are the solution to the differential equations.
\subsection{Sympolic PINNs}
\begin{figure}[t]
    \includegraphics[width = 0.47\textwidth]{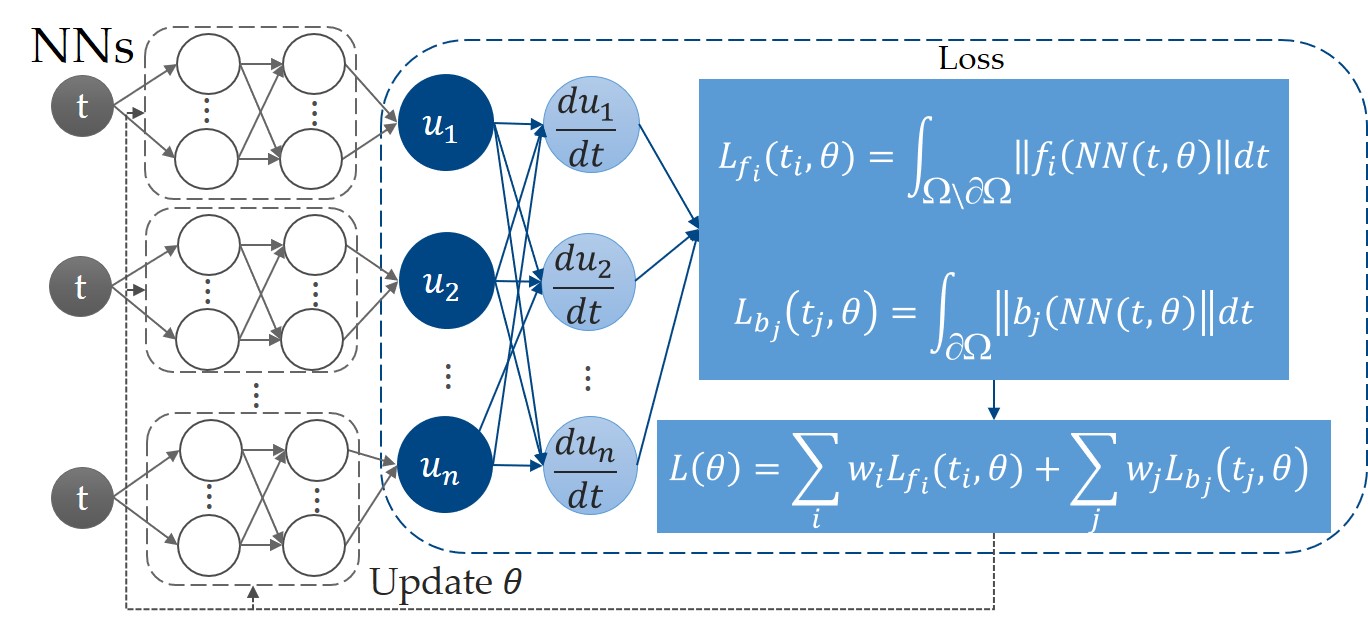}
    \caption{Schematic of symbolic PINNs.}
    \label{fig: symbolic_PINN}
    \vspace{-6pt}
\end{figure}
Suppose we leverage traditional PINNs to solve a set of $m$ equations with $n$ variables. The advantage of this method is that it enables larger Basic Linear Algebra Subprograms (BLAS) operations. 
However, if we want to split the equations into multiple single-output NNs, it can lead to performance issues with kernel saturation. 
In addition, when employing automatic differentiation (AD), both forward and reverse mode automatic differentiation scale with the number of outputs $\mathcal{O}(mn)$ and $\mathcal{O}(m+n)$, respectively. 
This means that if a differentiation term is only necessary for one variable, it can result in a significantly higher computational cost, as it will require computing the derivative with respect to all variables. 
Furthermore, this approach requires that all dependent variables be defined on the full domain, which is not always the case in PDEs \cite{zubov2021neuralpde}.

Using the symbolic technique allows us to represent the loss function in an abstract form while preserving its mathematical structure \cite{zubov2021neuralpde}. 
Hence, we can divide the loss function into components and integrate only the relevant domains for each portion.
In most cases, $f$ can consist of boundary conditions which are needed to be satisfied on (some subset) $\partial \Omega$. 
Therefore, $f$ can be separated into its component functions, and we obtain the following losses:
\begin{align}
     \mathcal{L}_{f_i}(t_i,\theta) = \int_{\Omega \backslash \partial \Omega} \| f_i(NN(t,\theta)) \|dt \label{residual_loss} \\ 
      \mathcal{L}_{b_i}(t_i,\theta) = \int_{ \partial \Omega} \| b_i(NN(t,\theta)) \|dt,
    \label{boundary_loss}
\end{align}
where $f_i$ is each equation in the system of PDEs, \eqref{residual_loss} represents the PDE loss term, $b_i$ are the boundary conditions and \eqref{boundary_loss} represents the boundary condition loss term.
The PDE loss term represents the residual obtained when substituting the outputs from the NN into the given PDEs, and the boundary condition loss term represents the difference between the outputs and the boundary conditions \cite{cai2021physics}.
To evaluate the PDE loss term, we need to find the set of points $t_i$.
A simple way to approximate the integral is to select a grid of $t_i$ and use the Trapezoidal method \cite{zubov2021neuralpde}:
\begin{equation}
    \int_{\Omega \backslash \partial \Omega} \| f_i(NN(t,\theta)) \|dt \approx \sum_i \Delta t \| f(t_i) \|.
    \label{Grid_method}
\end{equation}
Another method is to take $t_i$ at fixed numbers of random points and integrate them using Monte-Carlo methods \cite{zubov2021neuralpde}:
\begin{equation}
    \int_{\Omega \backslash \partial \Omega} \| f_i(NN(t,\theta)) \|dt \approx \alpha \sum_i \| f(t_i) \|,
\end{equation}
where $\alpha$ is the arbitrary constant.
In the end, each loss function of \eqref{residual_loss} and \eqref{boundary_loss} is multiplied by some factors and added together to form the final loss function:
\begin{equation}
    \mathcal{L}(\theta) = \sum_i w_i \mathcal{L}_{f_i}(t_i,\theta) + \sum_j w_j \mathcal{L}_{b_j}(t_j,\theta),
    \label{final_loss}
\end{equation}
where $w_i$ and $w_j$ are the factors associated with the PDE and boundary condition loss terms, respectively.
By calculating at multiple points in the domain, symbolic PINNs can avoid kernel saturation issues.
The structure of symbolic PINNs is shown in Figure \ref{fig: symbolic_PINN}.

\subsection{Adaptive weight of loss functions}
From \eqref{boundary_loss}, we can notice that the boundary conditions only computationally evaluate when $t \in \partial \Omega$ and would be 0 for all $t \in \Omega \backslash \partial \Omega$, causing dimensional of the boundary condition loss term is lower than the PDE loss term.
Commonly, NNs have low derivative frequency scale amplification in the component loss functions because the initial distribution of loss functions is fairly linear.
If the solution $u(t)$ has a high frequency in certain regions of the domain, the differences in scale of the component loss functions can become more pronounced and cause changes in relation to one another.
This is one of several ways the component loss functions have different scales, leading to many optimization difficulties during training the PINNs.
To address this obstacle, we need to adjust the gradients of the boundary condition loss term to the same scale as the gradients of the PDE loss term.
One method had been introduced in \cite{wang2020understanding}, in which an adaptive weight $\hat{w_i}$ can be defined as:
\begin{equation}
    \hat{w_i} = \frac{\max_\theta \{| \nabla_\theta \mathcal{L}_{f_i}(\theta_n) | \}}{\text{mean}_\theta\{ | \nabla_\theta \mathcal{L}_{b_j}(\theta_n) | \}}, \quad n \in S,
\end{equation}
where $\max_\theta \{| \nabla_\theta \mathcal{L}_{f_i}(\theta_n) | \}$ is the maximum of the absolute values of the gradients of the PDE loss terms, $mean_\theta\{ | \nabla_\theta \mathcal{L}_{b_j}(\theta_n) | \}$ is the mean of the absolute values of the gradients of the boundary condition loss terms, and $S$ is the set of certain iteration in which the adaptive weigh will be calculated (e.g., every iteration or every ten optimization iteration).
Then, the component factor associated with each loss term can be updated as follows:
\begin{equation}
    w_i = (1-\gamma)w_i + \gamma \hat{w}_i,
\end{equation}
where $\gamma$ is an additional hyperparameter.
The recommended value of $\gamma$ is 0.9 \cite{wang2020understanding}.

\section{Case studies and Implementation}
\label{casestudy}
\subsection{Physics model of power systems dynamics}
\label{physical_model}
\begin{figure}[t]
    \centering
    \includegraphics[width = 0.3\textwidth]{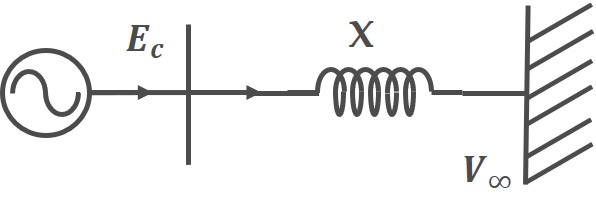}
    \vspace{-0.5em}
    \caption{Single machine infinite bus (SMIB) system \cite{mohammadian2023gradient}.}
    \label{fig:SMIB}
    \vspace{-5pt}
\end{figure}

In this paper, we consider a specific model of power system dynamics, which is a single-machine infinite bus (SMIB) system, shown in Figure \ref{fig:SMIB}, represented as follows \cite{sauer2017power}:
\begin{subnumcases}{\label{SMIB:ODEs}}
    \frac{d \delta}{dt} = \omega_t \\
    \frac{d \omega_t}{dt} = K_1 - K_2 \sin(\delta) - K_3 \omega_t,
\end{subnumcases}
\vspace{-\topsep}
\begin{equation*}
\text{where}~~~    K_1 = \frac{\omega_s}{2H}T_m, K_2 = \frac{\omega_s}{2H}\frac{E_cV_{\infty}}{X}, K_3 = \frac{\omega_s}{2H}D,
\end{equation*}
with $\delta$ is the rotor angle behind the transient reactance, $\omega_t = \omega - \omega_s$ is the transient speed ($\omega$ is the generator's angular speed).
The parameters used in \eqref{SMIB:ODEs} including $H$, the generator's inertia constant, $D$, the damping coefficient, $T_{m}$, the mechanical torque constant, $\omega_s$, the referenced angular speed, $E_c$ corresponding to the internal generator voltage, $V_\infty$ corresponding to the infinite bus voltage, and $X$ representing the sum of the generator's internal reactance and the reactance of the losses line.

\subsection{Implementation}
\label{implementation}
We perform the case studies in Julia Programming Language version 1.9.2 on a desktop - Intel core i9-10900.
To investigate the performance of symbolic PINNs, we consider the SMIB system in normal operation with $k_1 = 5$, $K_2 = 10$, and $K_3 = 1.7$ (the parameters can be found in \cite{sauer2017power} chapter 5.8), and the simulation duration is $[0, 10s]$.
The initial rotor angle, $\delta(0) = -1 rad$, and the initial transient speed, $\omega_t = 7 rad/s$. 
The implementation is the following steps:
\begin{itemize}[noitemsep,nolistsep, leftmargin=*]
    \item We implement the equation \eqref{SMIB:ODEs} and the initial conditions in Julia via the ModelingToolkit (MTK) package.

    \item We use the Lux package \cite{pal2023lux} to create an NN that solves the given equations. 
The NN architecture is built using the \textit{Chain} function, which has a fully connected layer defined by the \textit{Dense} function. 
The NN consists of two sub-networks that correspond to two state variables, $\delta$ and $\omega_t$. 
Each sub-network has an input dimension of 1, four hidden layers with ten neurons each, and an output dimension of 1.
The hidden layers use hyperbolic tangent as an activation function.

\item We leverage \textit{PhysicsInformedNN} function to present the NN as a trial solution. 
The outputs from the NN are substituted to the implemented equation and using \textit{GridTraining} function to estimate the loss function follow the method in \eqref{Grid_method} with a time steps of 0.01s.

\item By using a \textit{symbolic\_discretize} interface, we can take the loss function apart and then reassemble it before converting it into the optimization problem by the \textit{OptimizationProblem} function via the Optimization package and optimized by the Broyden-Fletcher-Goldfarb-Shanno (BFGS) optimizer \cite{wright2006numerical}.
\end{itemize}
We perform the study over 50 times, with the maximum iteration of each is 50,000.
To validate the results, we calculate the error between them with the results from the classical method by root means square method.
For the classical method, the equation \eqref{SMIB:ODEs} is solved by \textit{Tsit5} solver (similarly with \textit{ode45} solver in MATLAB) with a time step of 0.01s by using DifferentialEquations package \cite{rackauckas2017differentialequations}. 
After training, we evaluate the effectiveness of the transfer model of the previous case study with new initial conditions in two cases:
\textit{case 1}: initial angle $\delta(0) = 1\mbox{rad}$, and initial transient speed $\omega_t = -5\mbox{rad/s}$, \textit{case 2}: initial angle $\delta(0) = 0\mbox{rad}$, and initial transient speed $\omega_t = 2\mbox{rad/s}$.
Finally, to investigate the performance of the adaptive weight method, we consider the SMIB system in a faulty condition, namely pole slipping.
Pole slipping happens when the electromagnetic torque used to produce the power output is lower than the mechanical torque generated by a prime mover \cite{660906}.
Thus, we change a parameter $K_3$ to 1.6 while keeping the remaining parameters unchanged.
The adaptive weight is calculated every 10 iterations. 

\section{Numerical Results}
\label{Numerical}
\subsection{Performance of symbolic PINNs}
\label{pre-trained}
\begin{figure}[t]
    \centering
    \begin{subfigure}{0.48\textwidth}
        \begin{subfigure}{0.48\textwidth}
        \includegraphics[width=\textwidth]{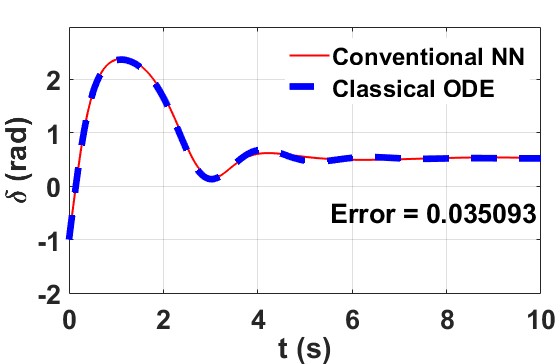}
        \end{subfigure} 
        \begin{subfigure}{0.48\textwidth}
        \includegraphics[width=\textwidth]{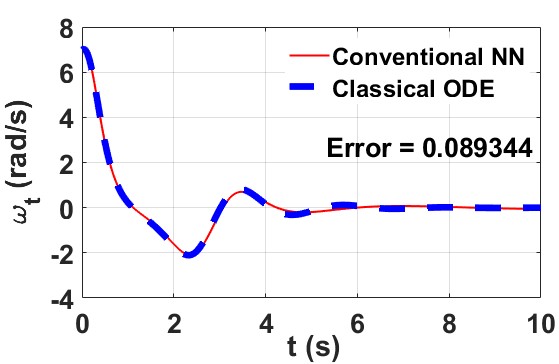}
        \end{subfigure}
        \vspace{-5pt}\caption{Conventional PINN}
        \label{fig: conventional}
    \end{subfigure}
    \begin{subfigure}{0.48\textwidth}
        \begin{subfigure}{0.48\textwidth}
        \includegraphics[width=\textwidth]{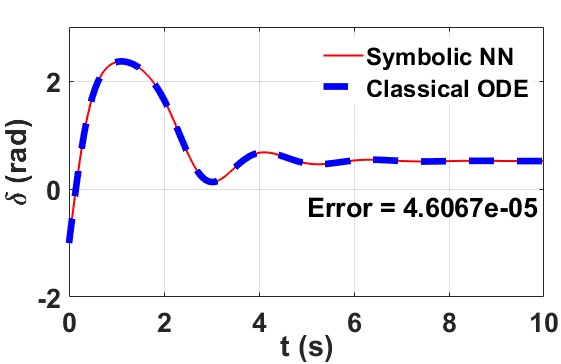}
        \end{subfigure} 
        \begin{subfigure}{0.48\textwidth}
        \includegraphics[width=\textwidth]{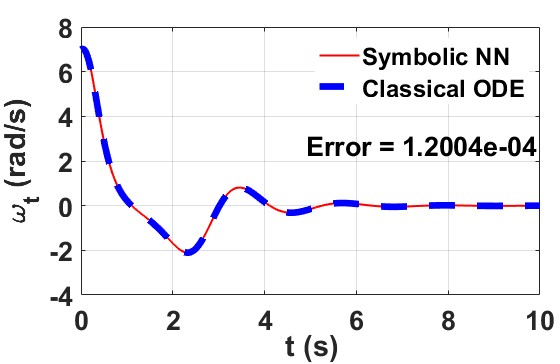}
        \end{subfigure}
        \vspace{-5pt}
        \caption{Symbolic PINN}
        \label{fig: symbolic}
    \end{subfigure}
    \vspace{-5pt}
    \caption{Results of SMIB of conventional and symbolic PINNs.}
    \label{fig: SMIB}
    \vspace{-5pt}
\end{figure}

Figure \ref{fig: SMIB} shows the results of the SMIB system (rotor angle, $\delta$, and generator's transient speed, $\omega_t$).
The dashed line represents the values from the classical method, the solid line represents the values from the PINNs, where Figure \ref{fig: conventional} illustrates the values of conventional PINN and Figure \ref{fig: symbolic} illustrates the value of symbolic PINN.
From Figure \ref{fig: conventional}, we can see that after training, the conventional PINN can model the power system dynamics with an accuracy upper 90\%.
However, the total parameters used in the conventional PINN are 1342 for 5 layers (including weights and biases), and the average time to train the network is 828s.
For the symbolic PINN in Figure \ref{fig: symbolic}, there is a total of 502 parameters for two sub-networks with 4 layers in each, the average training time is 332s, and the errors are very close to zero.
Based on the results, it is evident that the symbolic PINN provides much higher accuracy while using significantly fewer parameters and has a training time that is around 2.5 times faster than the conventional PINN.
The similar results can be found in \cite{sauer2017power}.

\subsection{Symbolic PINNs with transfer model}
\begin{figure}[t!]
    \centering
    \begin{subfigure}{0.24\textwidth}
        \begin{subfigure}{\textwidth}
        \includegraphics[width=\textwidth]{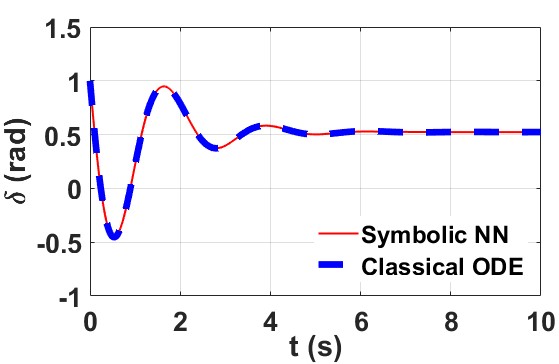}
        \end{subfigure} 
        \begin{subfigure}{\textwidth}
        \includegraphics[width=\textwidth]{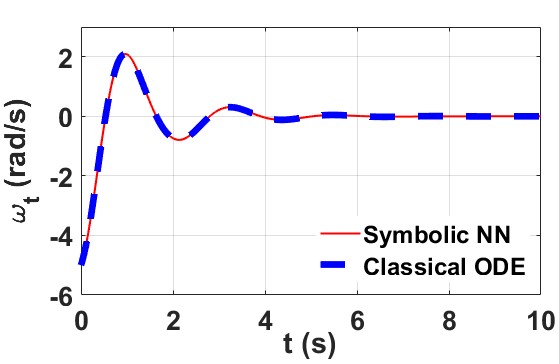}
        \end{subfigure}
        \begin{subfigure}{\textwidth}
        \includegraphics[width=\textwidth]{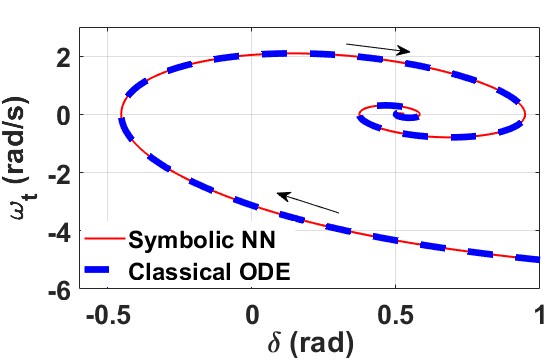}
        \end{subfigure}
        \vspace{-15pt}
        \caption{case 1}
        \label{fig: case1}
    \end{subfigure}
    \begin{subfigure}{0.24\textwidth}
        \begin{subfigure}{\textwidth}
        \includegraphics[width=\textwidth]{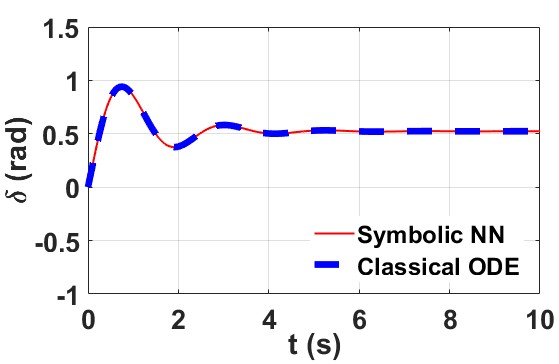}
        \end{subfigure} 
        \begin{subfigure}{\textwidth}
        \includegraphics[width=\textwidth]{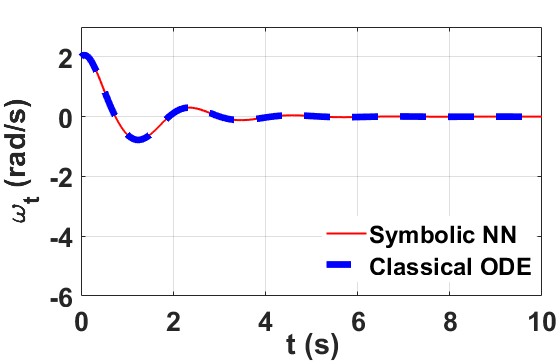}
        \end{subfigure}
        \begin{subfigure}{\textwidth}
        \includegraphics[width=\textwidth]{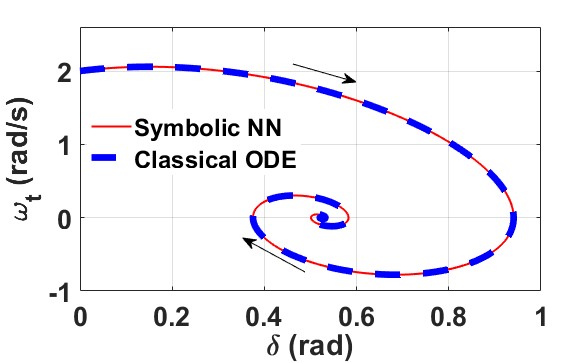}
        \end{subfigure}
        \vspace{-15pt}
        \caption{case 2}
        \label{fig: case2}
    \end{subfigure}
    \vspace{-17pt}
    \caption{Results of SMIB when changing initial conditions}
    \label{fig: SMIB_transfer}
    \vspace{-5pt}
\end{figure}
The training time of networks depends on the initial conditions. 
For the networks trained from scratch (without transfer parameters), we recorded the following results\footnote{Note: for the networks with transfer model, the optimal values are the same due to identity initial weights and biases, but have different training times because of the rate of the computer.}:
\textit{case 1}) the average training time is 363s and the average error is $2.85\cdot10^{-3}$, \textit{case 2}) the average training time is 295s and the average error is $8.15\cdot10^{-4}$.
For the network with the transfer model: \textit{case 1}) the average training time is 176s, and the error is $9.18\cdot10^{-5}$, \textit{case 2}) the average training time is 161s, and the error is $6.6\cdot10^{-5}$.
The results show that the transfer model reduces the training time by around two times and improves the networks' accuracy.
Additionally, the downside of the BFGS algorithm is that it may stop at the local optimal instead of the global optimal \cite{zubov2021neuralpde}.
With the transfer model, we can avoid getting stuck at the local optimal and reduce the total training time.

Figure \ref{fig: SMIB_transfer} shows the results from the symbolic PINNs using the transfer model after training in \ref{pre-trained}, where the first row represents the rotor angle, the second row represents the generator's transient speed, and the third row represents the phase portrait.
Phase portrait is the trajectories' generic representation of power system dynamics in the phase plane, which illustrates the behavior of the system by state variables.
The values of symbolic PINNs are represented in the solid line, and the values of the classical method are represented in the dashed line.
We can see that for different initial conditions, the trajectories of $\delta$ and $\omega_t$ are different, but the time to converge is nearly the same.
Furthermore, while the trajectories in phase portrait converged in different directions, they have the same stable point at (0.523, 0).

\subsection{Symbolic PINNs with adaptive weight}
\begin{figure}[t!]
    \centering
    \begin{subfigure}{0.24\textwidth}
        \begin{subfigure}{\textwidth}
        \includegraphics[width=\textwidth]{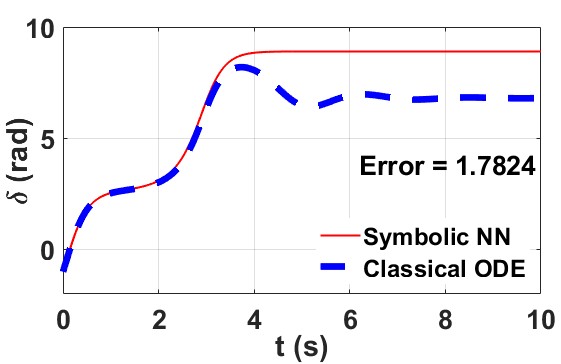}
        \end{subfigure} 
        \begin{subfigure}{\textwidth}
        \includegraphics[width=\textwidth]{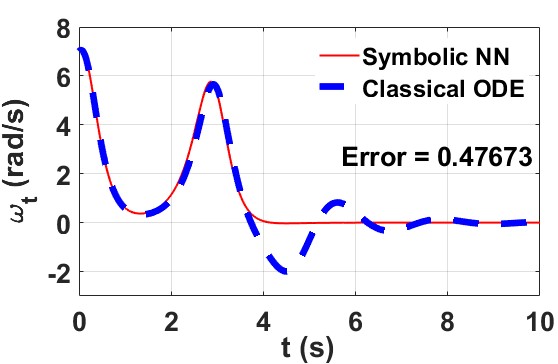}
        \end{subfigure}
        \begin{subfigure}{\textwidth}
        \includegraphics[width=\textwidth]{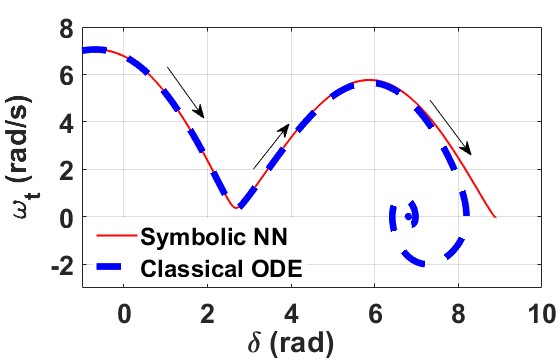}
        \end{subfigure}
        \vspace{-15pt}
        \caption{Without adaptive weight}
        \label{fig: non-adaptive}
    \end{subfigure}
    \begin{subfigure}{0.24\textwidth}
        \begin{subfigure}{\textwidth}
        \includegraphics[width=\textwidth]{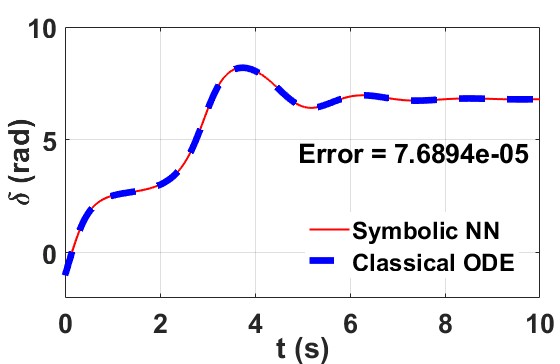}
        \end{subfigure} 
        \begin{subfigure}{\textwidth}
        \includegraphics[width=\textwidth]{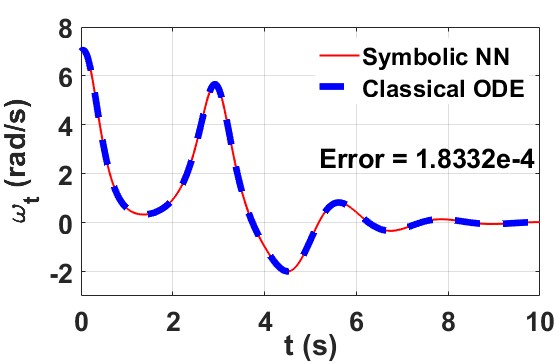}
        \end{subfigure}
        \begin{subfigure}{\textwidth}
        \includegraphics[width=\textwidth]{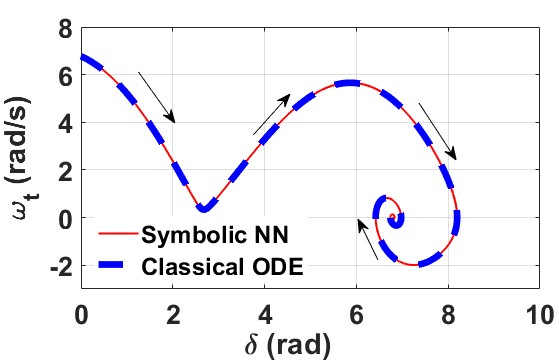}
        \end{subfigure}
        \vspace{-15pt}
        \caption{Having adaptive weight}
        \label{fig: adaptive}
    \end{subfigure}
    \vspace{-15pt}
    \caption{Results of SMIB in pole slipping}
    \label{fig: pole_slipping}
    \vspace{-10pt}
\end{figure}
Results of the SMIB system in pole slipping are shown in Figure \ref{fig: pole_slipping}, where Figure \ref{fig: non-adaptive} illustrates the values of symbolic PINNs without using adaptive weight, and Figure \ref{fig: adaptive} illustrates the values of symbolic PINNs with adaptive weight every 10 iterations.
The results shown in Figure \ref{fig: non-adaptive} is the best one, getting over 50 times.
The network can model the system in the first several seconds and then converge to the stable point, which is not the true equilibrium point.
The evidence for that is the rotor angle and transient speed reach a new stable point (first and second row of Figure \ref{fig: non-adaptive}) but in the phase portrait (third row of Figure \ref{fig: non-adaptive}) the trajectory of the values of networks did not converge to the same point as the classical one.
This is because the network got the vanishing gradient at the first time the transient speed $\omega_t$ lower than zero.
As a result, the errors of rotor angle $\delta$, and transient speed $\omega_t$ are 1.7824 and 0.47672, respectively.
After applying adaptive weight, the vanishing gradient issue was solved.
By reweighing the factors between loss function components, the system can be modeled well.
The accuracy of networks is nearly 100\% due to the errors of rotor angle and transient speed being very small ($7.6894\cdot10^{-5}$, and $1.8332\cdot10^{-4}$).

\section{Conclusion}
\label{conclu}
This paper represents a symbolic PINN model to simulate the power system dynamics and the adaptive weight method used to capture numerical instability.
We review the conventional PINNs and the physics model of power system dynamics, particularly the single-machine infinite buses (SMIB) system.
The model of the SMIB system is implemented into Julia by using the ModelingToolkit package, then using them as regularities in the PINNs.
After using the Lux package to create a network, the equation is discretized and trained using the PhysicsInformedNN function.
The loss function can be taken apart and reassembled by utilizing the symbolic\_discretize function in the NeuralPDE package, then embedded into the optimization problem by the Optimization package.
This loss function is minimized by the BFGS optimizer to find the optimized parameters.
Our results show that the symbolic PINNs can model the \color{black} SMIB system \color{black} with higher accuracy when they require fewer parameters and the training time faster than 2.5 times.
The transfer model helps us to reduce the time needed to train in the new initial conditions and avoid stuck at the local optimal when using the BFGS optimizer, with lower errors.
Finally, with the adaptive weight method, the networks would capture the optimization difficulties, such as vanishing gradient.
In the scope of the paper, we just focus on the SMIB system with two state variables, in the future, we will enhance the networks for the more complex systems.
\section{Acknowledgement}
This research is supported by the Alfred P. Sloan Foundation Grant \#10358 and the NCAT's Intel Foundation Gift.


\bibliographystyle{IEEEtran}
\bibliography{PINN}
\end{document}